# THE FORM OF GENTRIFICATION
## Common morphological patterns in five gentrified areas of London, UK.

A. Venerandi[1], M. Zanella[2], O. Romice[3], S. Porta[3].
[1]: Department of Civil, Environmental and Geomatic Engineering (CEGE), University College London, London, UK; [2]: Department of Mathematics and Computer Science, University of Ferrara, Ferrara, IT; [3]: Department of Architecture, University of Strathclyde, Glasgow, UK.

Abstract
Many socioeconomic studies have been carried out to explain the phenomenon of gentrification. Although results of these works shed light on the process around this phenomenon, a perspective which focuses on the relationship between city form and gentrification is still missing. With this paper we try to address this gap by studying and comparing, through classic methods of mathematical statistics, morphological features of five London gentrified neighbourhoods. Outcomes confirm that areas which have undergone gentrification display similar and recognizable morphological patterns in terms of urban type and geographical location of main and local roads as well as businesses. These initial results confirm findings from previous research in urban sociology, and highlight the role of urban form in contributing to shape dynamics of non-spatial nature in cities.

1. Introduction

Gentrification and more recently super-gentrification are well documented phenomenon in centres of financial dominance such as New York, Paris and London; however, they are happening at slower pace in many of our cities. Because urban fabric is complex, it is important to understand whether and to what extent urban form plays a role in initiating or supporting this process. Several theories/approaches have been developed to explain gentrification, its triggers, dynamics and protagonists. Generally these theories explain gentrification in terms of cultural, demographic, economic, and political changes; whilst very interesting observations on the quality of the place targeted or favoured by gentrifiers have often been recorded, urban form has rarely been at the forefront of this work. The urban form of gentrification is the focus of this paper. In particular, we are interested in understanding whether gentrification that starts from the bottom-up by collective activism does happen in urban areas that are physically and spatially similar and, if so, what are the physical and spatial features that these areas have in common.

The first part of the paper will offer an overview of the main aspects of gentrification – triggers, development, investment, protagonists and then move on to identify recurrent urban and architectural characteristics of gentrified areas. The central portion of this paper will then focus on the relationship between urban form and gentrification in five cases of gentrified urban areas in London, which have already been focus of studies on gentrification of a sociological nature.

*1.1. Overview of gentrification as a dynamic phenomenon*

Gentrification is a socio-spatial phenomenon that entails interlinked changes in the values of inner city areas, the degree of upgrade of both housing stock and services and the profile of their residents and visitors (Pacione, 2009).

Many have worked to define and describe gentrification firstly by identifying what triggers this process. S*upply-side theories* identify a late 1950s-early '60s capital investment as prime cause, and in particular the differential between the real value of property and the land value of inner city areas (N. Smith, 1987). The differential was due to a protracted lack of investment in inner city infrastructure following mid-high class suburbanisation; when the value of inner city stock became low enough to attract investment from developers or public agencies, to then re-sell at a profit, this paved the way for middle-income classes to return to centres, in search for more engaging ways of life than those found in the suburbs. This is to Smith a violent process of appropriation of value by middle classes. *Demand-side theories* (Ley, 1994), on the other hand attribute gentrification to the raise of the economic capacity and cultural profile of middle classes, following their transition from an industry-based economy to a service industry, and their dissatisfaction with contemporary urbanism and search for space with social meaning (Atkinson & Bridge, 2004; Lees, Slater, & Wyly, 2010).

Gentrification is a temporal phenomenon whereby socio, economic and cultural trends intervene, at different times, with nuances. Smith (2002) and Duany (2001), have identified 'waves' which generally occur in the process of gentrification in Europe and North America, the first of which took place from the beginning of the 1950s and saw a rather steady but sporadic migration into run-down areas by a cultural, artist-based elite. A second, more defined wave took, place in the 1970s-80s: here gentrification was linked to a process of economic restructuring through the migration of a higher, managerial and academic group into the same areas, encouraged by an improved overall quality carried out by piecemeal investment of the first wave of pioneers. Finally, a third, more generalised phase took place in the 1990s and over and saw the legal and financial sector moving in yet again in search of quality of both housing and services and therefore secure return for capital investment. All three phases imply in different degrees the transformation of inhabitants' profiles, in relation to their economic capacity to respond to raising cost of land and property values. Although views are contradicting, some believe that, as a process of urban change, gentrification is per se a positive one (Duany, 2001), granted that the transition between phases is at least generational, and not accelerated by artificial, large scale investment and population replacement, the "cataclysmic investment" discussed by Jane Jacobs (1961), allowing thus original lower income groups to benefit economically and socially from the progressive upgrade of their area.

Linked to the temporal dimension of gentrification, is therefore the issue of investment in the process: literature here contrasts investment by collective activism, to gentrification by capital. The former is closely associated with the first and in part the second temporal waves of the process, and the latter might be associated with the third (Butler, 2003). Investment in renewal is also, partly, associated with physical form: investment through collective activism relies more generally on small private capacity and therefore seems logic to expect it to favour piecemeal development, which is granted by a diverse, mixed form of buildings/units of different age, form and size. Capital investment – of corporate or public nature – is larger in scale and can therefore favour a less diverse, more unitary urban form.

The temporal dimension of gentrification may be associated with the cyclic waves of change over time of the built "fabric", as studied by urban morphologists since MRG Conzen's seminal work on Alnwick (1960). The "burgage cycle" is the gradual process of densification of plots through infill development of backyards that accompanies periods of socio-economic expansion, up to the clearing of buildings during stagnation that introduces to the initiation of the successive cycle (Slater, 1990). Though not linked with the process of gentrification as such, the notion of the "burgage cycle" introduces the idea that structural aspects of the urban fabric do change constantly end endlessly as a natural manifestation of urban life in time, an idea which is complementary to Duany's point about gentrification as a natural and cyclic dynamic of the urban historical process of change and renovation.

The physical dimension of places appears therefore a relevant factor in allowing the process of gentrification to start and develop in time, through a gradual progression which maintains a link between the socio-demographic profile, the economic capacity and the temporal transformation of an area whereby the original profile evolves rather than being substituted. The structural quality of the urban fabric appears to be a pre-requisite for gentrification to take place allowing existing population to benefit from a progressive process of socio, economic and environmental upgrade.

*1.2. Is there a recognisable form of gentrification?*

We intend to study the physical dimension of gentrification, to ascertain whether inner urban areas that have gone or are undergoing gentrification possess some physical shared property. If this is the case, it could have important implications on the control and management of processes of gentrification in our cities.

Several studies have identified that gentrified areas share a number of physical characteristics, such as the availability of substandard but structurally sound housing 'with potential', clustered to allow a contagion effect to occur, often with a unique amenity such as a view, proximity to good transport links with the central business district (CBD), and the presence of local commercial activities (shops, restaurants) attractive to gentrifiers (Pacione, 2009). Moreover, whilst housing is either generally gentrified in traditional, upgraded housing types or in converted industrial structures, the retail is generally gentrified in either piecemeal fashion, or through larger-scale interventions such as the 'festival marketplaces' i.e. Girardelli Square in San Francisco, Harbourplace in Baltimore, South Street Seaport in New York (Ellin, 1999).

From important recent studies, to which the core of this paper refers to, gentrified areas seem to possess well defined boundaries screening off less affluent areas (Butler, 2003), to contain well linked central spaces used as destinations with social character and use (Butler & Robson, 2001) mixed within areas with an ordered, pleasant and prosperous atmosphere, offering an overall sense of safety, and a family oriented feel where kids can have a degree of informally supervised independence (Butler & Lees, 2006). Streets are generally described as lined by terraced houses, not necessarily of any particular architectural merit, or by cottages and mews, and at times Victorian houses (Glass, 1964). Gentrified environments are often described as being dense and vibrant, with a good range of services accessible by foot, well connected to the centre whilst being not the centre themselves (Butler, 2003), but also conferring at the same time a sense of calm and order (Butler & Robson, 2001). Interestingly, the requirement for safety is linked to an open, interconnected urban form rather than to specifically designed features (i.e. these being specifically identified by respondents in interviews as the gated mid-rise block of the Chelsea, Hampstead and Notting Hill elite) (Butler, 2003).

*1.3. The structure of the paper*

This paper moves on from two previous studies that explore the sociological aspects of six neighbourhoods in London, which underwent a successful process of gentrification (Butler & Lees, 2006; Butler & Robson, 2001). Our aim is to explore the urban form of the same areas to understand a) whether common traits are recognizable in their morphological structure, and b) whether links are statistically recognizable between features of the urban form, and in particular between the centrality of streets and any of the other variables selected.

Our hypothesis is that there exists a recurrent city form that more than others is capable to respond positively to external (economic, cultural, social) pressures towards gentrification, by virtue of its inherent adaptability; a second hypothesis is that street centrality is functionally linked to some or all the structural aspects of urban form, and in particular is key to set the conditions for higher building density and closer built fronts. It should be noted, however, that the size of the cases investigated makes any generalization extremely adventurous and our conclusions should therefore be taken as tentative, with the value of the research being more on setting principles and methods than on providing conclusive evidence.

In section 2, we broadly present the historical, spatial and social reality of the cases studied, and introduce the indices utilized in this research to measure their form and the statistical methods deployed for their analysis. In section 3 we separately offer the research's findings for the analysis of the street network and the structure of the urban fabric. Finally, in section 4 we draw our conclusions.

## 2. Method

### 2.1. Case studies

In this section we illustrate how we selected the case studies and provide a basic description of their character.

#### 2.1.1. Selection of cases.

The selection of the study areas is based on Butler and Robson's "Social Capital, Gentrification and Neighbourhood Change in London: A Comparison of Three South London Neighbourhoods" (Butler & Robson, 2001). In this important work the authors identify 8 areas in London that underwent a clear trajectory of gentrification, to then focus their attention on three of them: Telegraph Hill, Battersea and Brixton. In our study we pick up these latter three, for which the authors provide detailed geographical boundaries that we assume with no modifications, and also two of the others for which we could identify clear boundaries ourselves: Barnsbury and Dalston. Case studies are visible in Fig. 1.

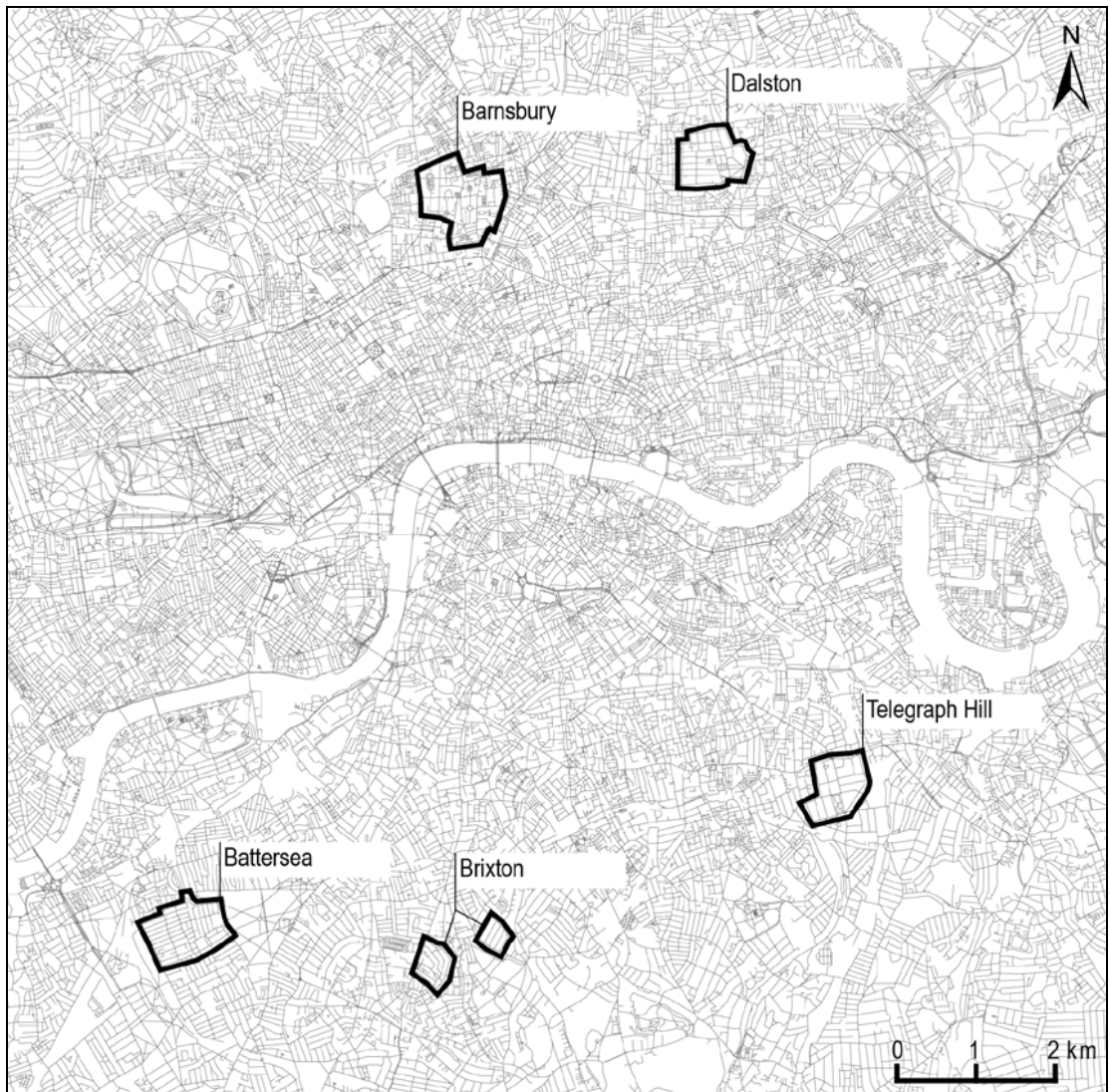
Fig. 1 Location of the five case studies in the 15km x 15km street network map of central London.

All the resulting five cases of this study have undergone processes of gentrification by collective activism, as defined in the Introduction above; four out of five are located between 3 and 4 miles from the city centre, with only Barnsbury sitting closer to it (about 2 miles).

2.1.2. Description of cases

*Barnsbury (Ba)*

Barnsbury is located in the Borough of Islington in North London at about 2 miles from the city centre. It was built at the beginning of the XIX century to host upper middle-class families. It is characterized by a mixture of beautiful detached villas and terraced houses. Initially inhabited by upper-income professionals, it soon became a middle-class neighbourhood and between the Two World Wars it was mainly inhabited by the working class. Gentrification started in the 1950s and reached its peak between the 1960s and 1970s. Today Barnsbury can be considered fully gentrified (Carpenter & Lees, 1995, p. 3-4).

*Battersea (Bt)*

This neighbourhood is located in the Borough of Wandsworth in South-West London at about 4 miles from the city centre. It underwent a process of urban regeneration in the last 20 years which improved its overall conditions. Battersea is characterized by semidetached Victorian houses and is mostly populated by the 'upper' middle class. Numerous wine bars and bistros populate Northcote Road, the main street of the neighbourhood. Two big parks (Wandsworth Common and Clapham Common) are located on the East and on the West side of the neighbourhood – hence its commercial nickname as "Between the Commons" (Butler & Robson, 2001, p. from 8 to 10).

*Brixton (Br)*

Brixton is a neighbourhood located in the borough of Lambeth in South London at about 3.5 miles from the city centre. It is characterized by a strong cultural and ethnical heterogeneity. Many waves of gentrification occurred in the area, the most important of which took place during the 1980s and has continued till now. It is characterized by semidetached Victorian and Georgian houses and it is mainly inhabited by middle class population who although have similar income, exhibit different attitudes in terms of social engagement (Butler & Robson, 2001, p. 12-13). The substantial Brockwell Park, is located in the southern part of the neighbourhood.

*Dalston (Da)*

Dalston is located in North-East London in the Borough of Hackney at about 3 miles from the city centre. It is mainly characterized by Georgian houses and it is well known for the park (London Fields) which lays on the eastern part of the neighbourhood. This park hosts several amenities - a public swimming pool, a pub and a café - which attract many people, especially during the weekends. Gentrification processes increased in the last years after the completion of a new railway station at Dalston Junction, part of the extension of the East London Line finished in 2010 and mainly due to the revitalisation of East London for the 2012 Olympics. According to census data, today more than 50% of its residents are in managerial, administrative and professional positions (LBH Policy and Partnership, 2013, p. 12).

*Telegraph Hill (Th)*

This neighbourhood is located in the Borough of Lewisham in South-East London at about 4 miles from the city centre. Four streets that slope up toward Telegraph Hill Park constitute its main core, and it is mainly characterized by Victorian houses. The turning point for its regeneration from a derelict to a well off area occurred in the early 1980s. The process lasted a decade and now the area is mainly characterized by the presence of middle-class people with families. The neighbourhood is a calm and quiet residential island with almost no commercial amenities: the only shop is a big supermarket located in the northern part of the study area (Butler & Robson, 2001, p. 5-6).

## 2.2. Variables and definitions

In this section we present the spatial elements that we use to represent the form of the city, and we also clarify how we measure them. First of all, we distinguish between the street network and the urban fabric.

## 2.2.1. Measuring the street network

Indices of the street network are derived from the physics of complex networks and particularly from the recent wealth of studies in networks of a spatial nature (Barthélemy, 2011). The importance of the connectivity of the street network to understand and tackle fundamental urban dynamics such as location of land uses, safety, navigability and prosperity, beyond the mere notion of accessibility, has been firstly recognized by Hillier and Hanson (Hillier & Hanson, 1984) to gain momentum among urban scholars in the past decade (Stangl & Guinn, 2011) and start to inform the planning agenda of global decision makers (U.S. Green Building Council, 2009; UN-Habitat, 2013a, 2013b; WA Ministry for Planning, 2009). Among the properties of the street networks that are relevant for urban life and functioning, street centrality certainly holds a special importance (Porta et al., 2013; Porta, Latora, & Strano, 2010). We dedicate the next paragraph to the presentation of the index utilized for the street network analysis of the five gentrified areas in London.

For the purpose of our study, we select a portion of the London's street network defined by a square of 15km of edge, centred in London's city centre (Fig. 1). In this network, one "street" is identified by its centreline from intersection to intersection, i.e. by a "primal" representation. The 15kmx15km extract is large enough to include the five study areas in one single fully interconnected network and yet to ensure that a sufficient buffer is left to avoid the insurgence of the edge effect, i.e. the distortion of centrality values that occurs at the external fringe of any study area just because of the proximity to the cut.

Over this 15kmx15km network, we map the centrality of streets utilizing the Multiple Centrality Assessment (MCA) approach. MCA was firstly introduced by Porta et al in 2006; since then it has been widely utilized to map street centrality according various indices such as Efficiency, Betweenness, Closeness, Straightness, or Information (Porta, Crucitti, & Latora, 2006). In this study, we only use the index of Betweenness centrality $C^B$, defined below, to characterize the links of the graph. $C^B$ is calculated firstly on the nodes of the graph, and then averaged on the links which connect them. For brevity and clarity, we will refer to $C^B$ as *Centrality* in the rest of the paper. *Centrality* on links (streets) constitutes the database that is processed in the analysis of correlation and variance in section 3. Differently, the visual analysis of centrality reported in section 3.1.2 is based on the Kernel Density Estimation (KDE) of *Centrality*. This method offers a number of substantive and visual advantages over the simple analysis of centrality on the network (Porta et al., 2009). KDE is a method of spatial smoothing which estimates the density of events within a range (circular window) from an observation point, to represent the value at that point. Within that window, KDE weighs nearby events more than distant ones, based on a *kernel function* (Bailey & Gatrell, 1995; Fotheringham, Brunsdon, & Charlton, 2000; Silverman, 1986). By distributing the observation points along a grid that covers the whole study area, and setting a certain pace for the grid (cell edge *c*) and a certain radius for the window (bandwidth *h*), we can generate a density of events (discrete points) as a continuous field (for example a raster). In this study, we set a cell edge *c*=10mts and a bandwidth *h*=100mts.

*Centrality* is based on the idea that a street is more central when it is "passed by" by a larger number of the shortest paths connecting each street to each other. It captures a very peculiar – and relevant – way of "being central" for a place, that of "staying in between" places along the most convenient routes that link any couple of them. If $n_{jk}$ is the number of shortest paths connecting two nodes *j* and *k*, and $n_{jk}$ is the number of shortest paths connecting two nodes *j* and *k* that contains node *i*, *Centrality* for node *i* is defined as (Freeman, 2013):

$$C_i^B = \frac{1}{(N-1)(N-2)} \sum_{j,k \in \mathbb{N}, j \neq k, k \neq i} \frac{n_{jk}(i)}{n_{jk}}$$

*Centrality* can have values between 0 and 1 and reaches its maximum when node *i* lies on all the possible shortest paths. It is demonstrated that Betweenness centrality nicely captures several key-features of the urban environment, for example population and employment density (Wang, Antipova, & Porta, 2011), presence of retail and services (Sergio Porta et al., 2009; Produit et al., 2010) and location of historical paths that shape the evolution of cities (Strano, Nicosia, Latora, Porta, & Barthélemy, 2012).

2.2.2. Measuring the urban fabric

Measures of the urban fabric refer to urban morphology as a separate area of urban studies. This area emerged in its current form mostly from two different "schools" in the late 50s and 60s of the XXth Century: the Italian strand that originated from Saverio Muratori and was then developed by G Caniggia, GL Maffei and others (Cataldi, Maffei, & Vaccaro, 2002), and the British strand that originated from the German-born MRG Conzen, then developed by the Birmingham group of JWR Whitehand, TR Slater and others (Whitehand, 2001). Nowadays Urban morphology is an expanding area of knowledge with very significant contributions form all over the world. While the quantitative measurement of structural features of urban form has always been part of the disciplinary toolbox in modern urban morphology, and although the quantitative approach has clearly gained momentum with the internationalization of the field since the late Eighties of the XXth Century under the notion of "metrological analysis" (Song & Knaap, 2004), the search for a systematic code of practice for the consistent measurement and statistical interpretation of the ordinary fabric of cities has yet a long way ahead. This paper is a contribution towards the progress of such systematic code of practice, or a science of urban morphometrics (Porta et al., n.d.). In section 2.2.2.1 we discuss the unit of analysis that sits at the heart of our study, the "street edge", while in section 2.2.2.2 we offer a definition of the eight indices utilized in this work.

2.2.2.1. The unit of analysis: the street edge

Quantitative explorations of the form of the urban fabric have utilized in the recent past a range of units of analysis at various scales, from the entire city to the district, the block, the plot, the building (or building type) within the plot, and even single buildings components within the buildings, spanning from macro to micro-morphology. A particular emphasis has been posed since the early 80s of the XXth Century to the urban block, interpreted as the building component of the compact-dense city of a pre-modern era in close link with theories of sustainable urbanism and place-making after the end of modernism (Power & Burdett, 1999). However, from an urban morphology point of view, designing blocks is not enough, nor really the point. Panerai et al (Panerai, Castex, & Depaule, 2004) strongly argue against the simplistic approach of too many urban designers who too often ignore the importance of the inner structure of the block, an attitude that is ultimately conducive to fake cities made of "pseudo-blocks". According to Caniggia and Maffei (Caniggia & Maffei, 2001) the process of formation of the traditional city does not proceed block by block, but rather through the piecemeal densification of streets edges. This is a plot by plot process that moves forward from the most central to the least central streets, ultimately leading to the generation of blocks. As a result, blocks have always been the outcome of a historical formation whose basic unit of development is the plot, and more appropriately the street edge, defined as the combination of plots fronting the same street.

According to this understanding, we define "street edge" of the Block *b* on the street *s* the portion of land $E_{bs}$ that is constituted by all the plots that have their main access from *s*.

The consequences of this kind of process appear to significantly drive the scale of the traditional city with remarkable impacts on a number of important aspects of city life and functioning (Mehaffy, Porta, Rofè, & Salingaros, 2010; S. Porta, Romice, Maxwell, Russell, & Baird, 2014). One of the most important implications of this fine-grained, plot-based, street edge by street edge development, is the structural complexity of the resulting urban block, which greatly enhances robustness of the urban fabric, i.e. its ability to successfully accommodate changing circumstances, users and needs in time. Gentrification, especially when happening by collective activism like in the five cases selected for this study, typically is the manifestation of such a social and economic pressure for change. We therefore utilize the Street edge as unit of analysis – rather than the block – as reasonable way to explore urban areas that have undergone such a process of change and proved a significant ability to do that without major cataclysmic disruptions to the physical form of the place itself. In the next paragraph we present a clear definition of the street edge as well as one of each of the indices used in the analysis of the urban fabric.

### *2.2.2.2. Indices of urban fabric*

It is a specific methodological choice of this research to rely only upon information that can be gathered remotely, taking advantage of repositories that are: a) publicly accessible, and b) capable to deliver information of the same quality all over the UK. The rationale for this is that we want to introduce and test a method that could ideally be extended to support a much wider effort in the future, potentially leading to the creation of an Atlas of urban form at the national scale. Source of information for the analysis of the urban fabric in this paper is the Ordnance Survey (OS) Master Map, which offers individual buildings, properties lines and other features in vectorial format. On occasions, OS information is integrated with insights from Open Street Map. We also regularly recur to Google Map and Street View for an in-depth inspection of the cases to resolve issues of interpretation of information from the previous sources, and to test ideas. The variables that we take into account in the analysis of the urban fabric are listed in Tab. 1 and exemplified graphically in Fig. 2.

| Code | Name | Definition | Unit | Range | Formula |
|------|------|------------|------|-------|---------|
| M1 | Street edge | The area of the street edge | m² | -- | -- |
| M2 | Coverage | The total area of the buildings' footprints | m² | -- | $M_2 = \frac{1}{n}\sum_{i=1}^{n} nB_i$ |
| M3 | Density | Total amount of gross floor area over the street edge area | m²/m² | -- | $M_3 = \frac{M_2 \times M_7}{M_1}$ |
| M4 | Street width | The prevalent street width between the two sidewalk lines | m | -- | -- |
| M5 | Coverage ratio | % of land covered by buildings on the street edge area | m²/m² | 0 - 100 | $M_5 = \frac{M_2}{M_1} \times 100$ |
| M6 | Centrality | $C^B$ of the street that serves the street edge | -- | -- | *Def. in 2.2.1* |
| M7 | Front height | Average height of all buildings in a street edge | No. of floors | -- | $M_7 = \frac{1}{n}\sum_{i=1}^{n} h_i$ |
| M8 | Built front ratio | % of the street edge front lined up by buildings | m/m | 0 - 100 | $M_8 = \frac{BF}{F} \times 100$ |

Table 1. List of the indices of urban fabric. In M2: *n*=number of buildings in the Street edge; $B_i$=Footprint area of building *i*. In M7: *n*=number of buildings in the Street edge; *h*=height (in number of floors) of building *i*. In M8: *BF*=length of the portion of the Street edge front that is lined by buildings within 8m from the sidewalk line; *F*=length of the Street edge front.

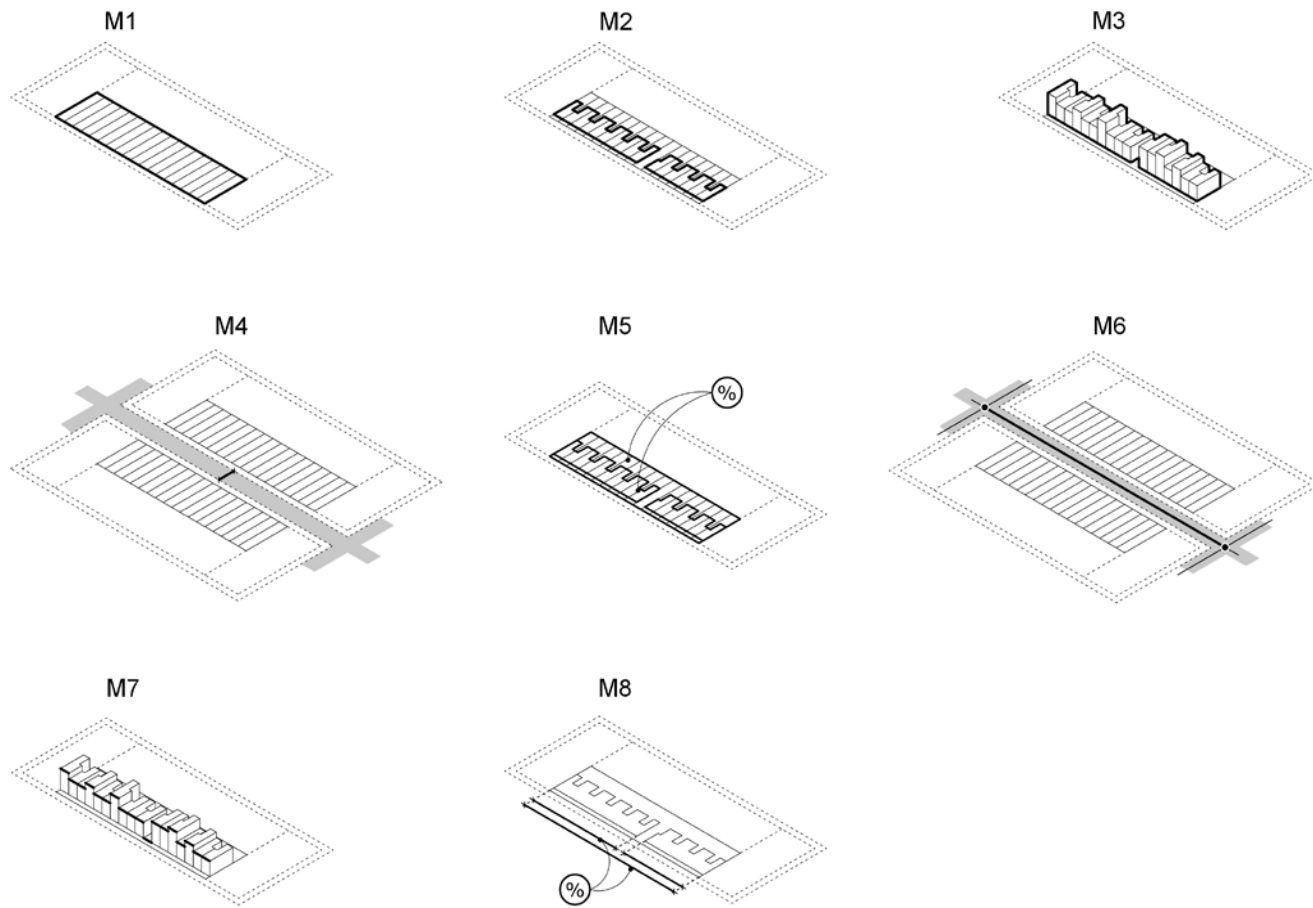

Fig. 2. Illustrational sketches of the indices of urban fabric from M1 to M8.

*2.3. Statistical analysis*

We performed statistical analysis taking into account the five neighbourhoods of London introduced above: Battersea, Barnsbury, Brixton, Dalston and Telegraph Hill. Each of these is described through the totality of its blocks, each of them studied as a combination of street edges, according to the definition offered in section 2.2.2.1. This technique allows us to obtain a large amount of data, for each neighbourhood, necessary condition for applying the Central Limit Theorem (CLT). This approach represents a very fundamental result in Mathematical Statistics about the convergence of a sum of $n$ independent and identically distributed random variables, with mean $\mu$ and finite variance $\sigma^2$, to a Normal Distribution $N(n\mu, n\sigma^2)$. The (CLT) is a precondition for obtaining the consistency of the statistical analysis.

Our work in this section takes into account several well-known statistical tests such us: Analysis of Variance (ANOVA), Linear Regression (LR) models in order to understand the emergence of common patterns with respect to the different neighbourhoods, and to estimate the relationships among the variables. Moreover we studied the statistical correlations through the computation of the Pearson Coefficients (PC), which identify the correlations based on the definition of Covariance of a set of random variables. The only non-parametric test considered in this work is the Kolmogorov-Smirnov Test (KST).

An introduction to the mathematics behind the cited methods can be found in classical books of Mathematical Statistics and estimation theory such us (Roussas, G., 1997; Fisz, M., 1962) and (Devroye, L., 1987).

3. Results

*3.1. Analysis of the street network*

We present in this section the results obtained from the analysis of the street network for the five case studies and we explain them through a visual analysis.

We present in Fig. 3 five extracts – one for each case study – of the centrality analysis performed on the 15kmx15km primal graph of London's street network introduced in 2.2.1. These maps represent the Kernel density of $C^B$ calculated on nodes and then averaged on links as discussed in 2.2.1.

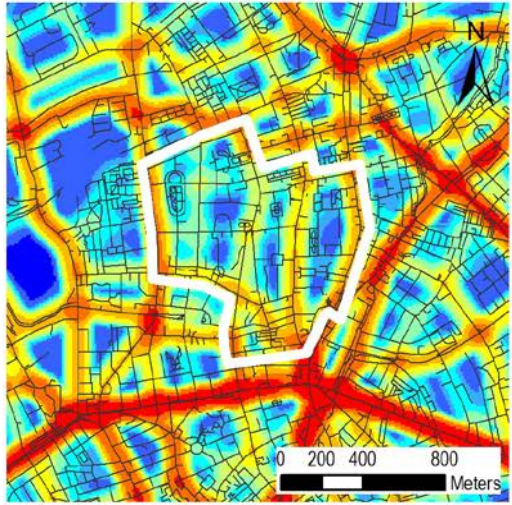
Barnsbury

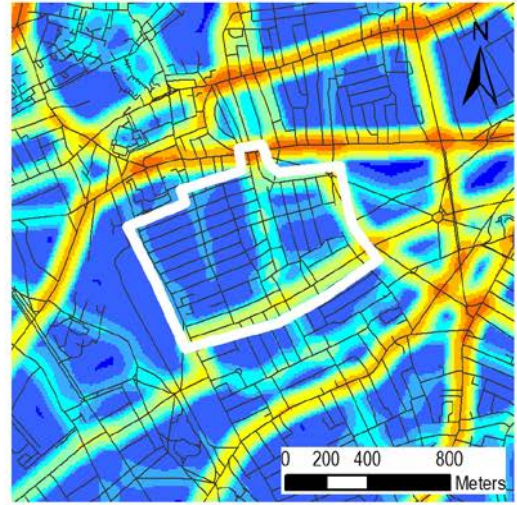
Battersea

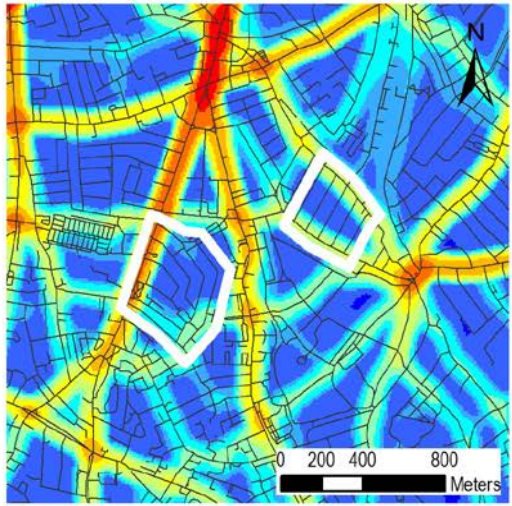
Brixton

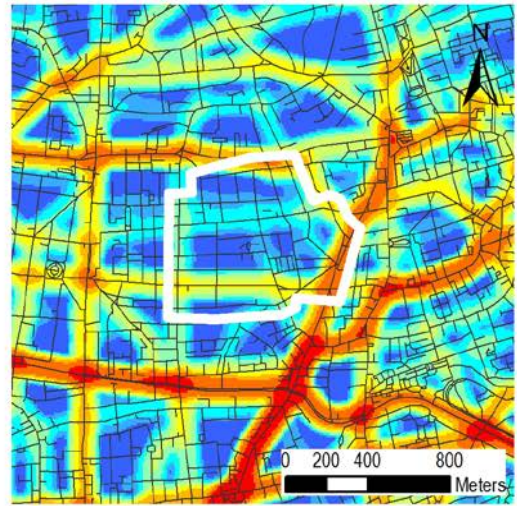
Dalston

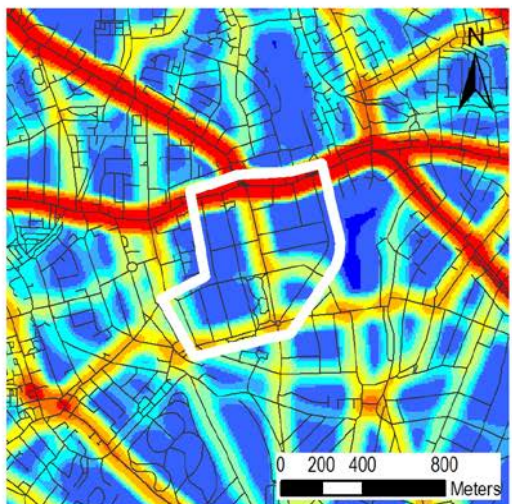
Telegraph Hill

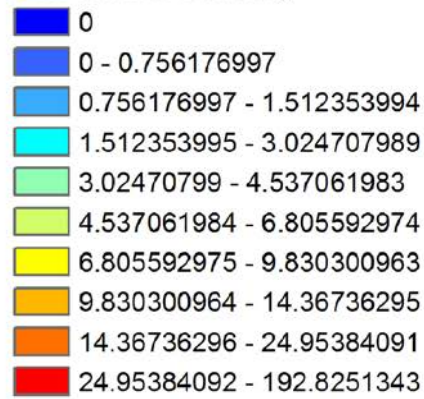

Fig. 3. Kernel density of centrality for the five case studies.

Visual inspection of the extracts highlights one common pattern: highly central streets ("urban roads", tending to red) do not cut the case areas, but rather lie on their boundaries. For example, we notice that the highest centrality streets of Barnsbury are located at its West (Caledonian Road), East (Upper St.) and South (Pentonville Road) boundaries. Central streets of a second grade ("local mains", tending to yellow) penetrate well within the case areas splitting them in two. This double system of main streets frames a "background" system of low centrality spaces ("locals", tending to blue). Since centrality creates the potential for intense urban activity, we argue that gentrified neighbourhoods tend to be calm, safe and mainly residential areas in their cores, connected to vibrant and busy roads at their edges by a system of intermediate locally central streets.

*3.2. Analysis of the urban fabric*

3.2.1. Analysis of Variance (ANOVA)

Overall the urban form of the five selected case studies exhibits, with some exceptions worth noting, remarkable similarities for all variables.

The *Street edge (M1)* comprised within the Interquartile Range (IQR) of each neighbourhood roughly ranges between 1,000 and 10,000m$^2$, the means falling in the interval of 4,000-5,000m$^2$ with only TH above 5,000m$^2$. Brixton offers a good representation of the centre-scale of values in the selected group of neighbourhoods, with a mean sitting around 4,000m$^2$, and an IQR ranging between 2,000-6,000m$^2$. Ba and Bt sit at the bottom of the scale; we notice that their means are statistically comparable and their values can be considered as equal in the context of the hypothesis tests. To support this thesis we present Tab. 1a, which shows the ANOVA outputs for Ba and Bt. In this table: the number of observations corresponds to the number of street edges of the two cases taken together; *Degrees of freedom (DF)* are the number of free components in a linear model, the F-test is a statistical test used to ensure the best fit of data via the computation of the Fisher-Snedecor distribution, while in the last column we report the P-value. This latter element is also called significance level of a test and is related to test statistics. Again a key reference for these topics is (Roussas, G. 1997). In Tab. 1a, we also present: means, standard deviations, lower observation and higher observation for the two mentioned neighbourhoods.

| Neighbourhoods | Nr. of Obs. | DF | F | Pr>F |
|---|---|---|---|---|
| Ba-Bt | 427 | 1 | 3.67 | 0.0562 |

| Neighbourhood | Nr. of Obs. | Mean | Std. Dev. | Min | Max |
|---|---|---|---|---|---|
| Barnsbury | 230 | 1616.23 | 2241.41 | 0 | 18210.17 |
| Battersea | 197 | 2008.07 | 1933.25 | 0 | 12035.81 |

Tab. 1a. ANOVA outputs and summary statistics for Ba and Bt.

In Tab.1a, we notice that the latter element *Pr>F* takes values 0.0562 which means that, taking a level of significance of the 95%, the means of the neighbourhoods Ba and Bt, respectively $\mu_{Ba}$ and $\mu_{Bt}$, pass the hypothesis test

$$H_0 : \mu_{Ba} = \mu_{Bt.}$$

In other words it is statistically admissible the statement that the two measures are equal. The same observation, with an even stronger level of evidence, can be made for Br and Da. We present in Tab. 1b ANOVA outputs and summary statistics for the two neighbourhoods.

| Neighbourhoods | Nr. of Obs. | DF | F | Pr>F |
|---|---|---|---|---|
| Br-Da | 159 | 1 | 0.05 | 0.8275 |

| Neighbourhood | Nr. of Obs. | Mean | Std. Dev. | Min | Max |
|---|---|---|---|---|---|
| Brixton | 49 | 3678.82 | 2894.52 | 0 | 10704.90 |
| Dalston | 110 | 3567.75 | 2962.52 | 0 | 14038.60 |

Tab. 1b. ANOVA outputs and summary statistics for Br and Da.

Although statistically equal only for the two pairs above, values of M1 are not significantly dissimilar across the five case studies. We report in Tab. 1c the percentiles Q1 (25%), Q2 (50%), Q3 (75%) for each neighbourhood. Q1 expresses the area of the largest Street edge in the group of the first 25% of all observations, ranked from the smallest to the largest, and analogously for Q2 (for the first 50%) and Q3 (for the first 75%).

| Neighbourhood | Q1 | Q2 | Q3 |
|---|---|---|---|
| Barnsbury | 0 | 941.54 | 2226.73 |
| Battersea | 746.72 | 1409.30 | 2968.90 |
| Brixton | 1132.74 | 2960.39 | 6075.77 |
| Dalston | 1540.28 | 2969.92 | 4445.41 |
| Telegraph Hill | 1045.97 | 5343.78 | 9975.11 |

Tab. 1c. Quartile intervals of M1 for the five case studies.

We conclude the analysis by presenting distributions of M1 for each neighbourhood (Fig. 4). In Fig. 4., we present two information: a) bar chart: the width of each bar varies according to a certain interval of values (in this case street edges between a certain area $a$ and $a + 2,000m^2$), while the height varies according to the percentage of observations falling within that interval; b) red line: the curve that a normal distribution would take assuming the estimated mean and variance for the actual dataset.

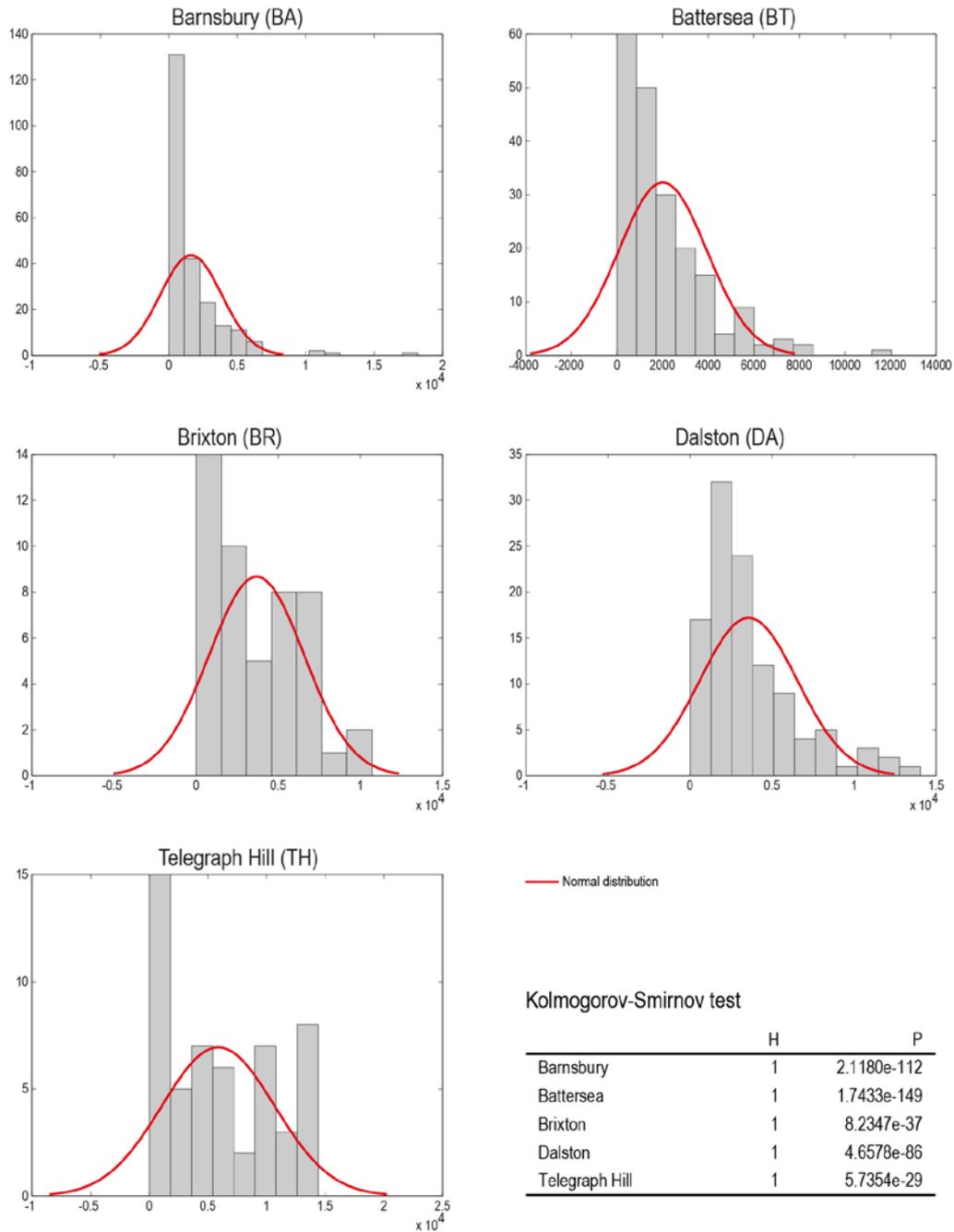

Fig. 4. Distributions of *Street edge (M1)* for the five neighbourhoods.

Distributions of M1 do not appear to follow a normal distribution pattern. Two pairs of cases (Ba-Bt and Br-Da) have similar mean values but their distributions are dissimilar.

*Coverage (M2)* ranges within a contained span of values, between 1,000-2,000m², and follows a function that is similar to the one found for the *Street edge (M1)*. The mean values for the variable M2 are comparable in the same vein of M1. This result is coherent with what we will illustrate later in this section; that is variables M1 and M2 are strongly linearly correlated. Tab. 2a makes the case of Ba and Bt:

| District | Nr. of Obs. | DF | F | Pr>F |
|---|---|---|---|---|
| Ba-Bt | 427 | 1 | 2.73 | 0.0995 |

| Neighbourhood | Nr. of Obs. | Mean | Std. Dev. | Min | Max |
|---|---|---|---|---|---|
| Barnsbury | 230 | 756.5598696 | 1309.38 | 0 | 11511.68 |
| Battersea | 197 | 932.4095408 | 772.7814335 | 0 | 3473.03 |

Tab. 2a. ANOVA outputs and summary statistics for Ba and Bt.

While in Tab. 2b we present results for Br and Da.

| Districts | Nr. of Obs. | DF | F | Pr>F |
|---|---|---|---|---|
| Br-Da | 159 | 1 | 2.57 | 0.1109 |

| Neighbourhood | Nr. of Obs. | Mean | Std. Dev. | Min | Max |
|---|---|---|---|---|---|
| Brixton | 49 | 1380.75 | 1131.44 | 0 | 4773.05 |
| Dalston | 110 | 1125.53 | 812.3234478 | 0 | 3990.63 |

Tab. 2b. ANOVA outputs and summary statistics for Br and Da.

We note that M2 and M1 have similar statistical behaviors. We conclude the analysis reporting quartile intervals of M2 (Tab. 2c), and M2 distributions for the five case studies (Fig. 5).

| Neighbourhood | Q1 | Q2 | Q3 |
|---|---|---|---|
| Barnsbury | 0 | 475.145 | 970.720 |
| Battersea | 343.905 | 733.985 | 1342.855 |
| Brixton | 538.93 | 1048.18 | 2213.62 |
| Dalston | 612.930 | 935.090 | 1472.260 |
| Telegraph Hill | 316.71 | 1458.86 | 2773.67 |

Tab. 2c. Quartile intervals of M2 for the five case studies.

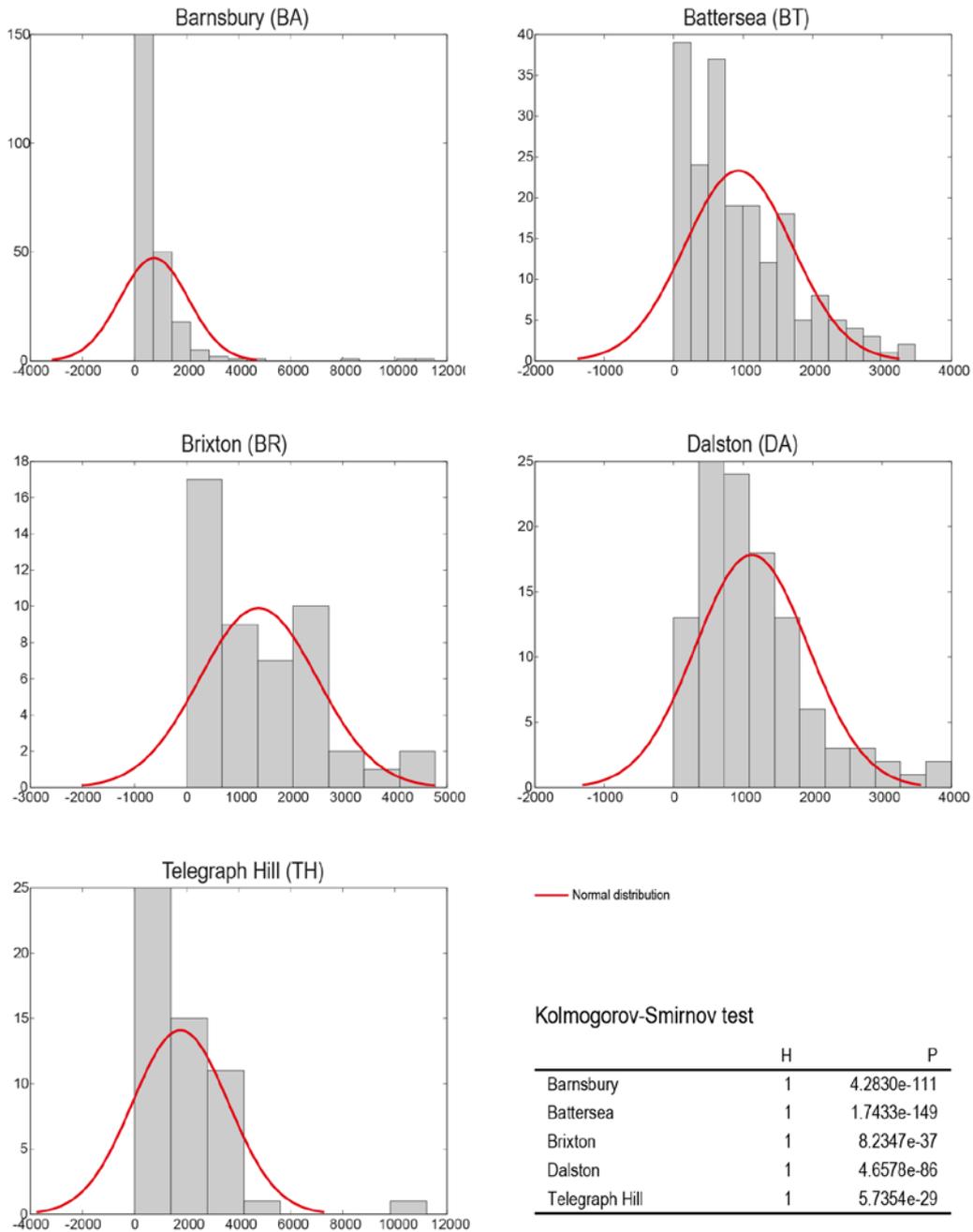

Fig. 5. Distributions of *Coverage (M2)* for the five neighbourhoods.

*Coverage ratio (M5)* shows for all neighbourhoods a quite contained set of values, with all means ranging between the 30-50% of the street edge areas. It shall be noted that for Ba, Br and Da the ANOVA procedure for testing

$$H_0 : \mu_{Ba} = \mu_{Br} = \mu_{Da}$$

allows us to accept the null hypothesis, which confirms that the three means are statistically equal. ANOVA outputs and summary statistics for M5 are showed in Tab. 3a.

| Districts | Nr. of Obs. | DF | F | Pr>F |
|---|---|---|---|---|
| Ba-Br-Da | 389 | 2 | 0.96 | 0.3819 |

| Neighbourhood | Nr. of Obs. | Mean | Std. Dev. | Min | Max |
|---|---|---|---|---|---|
| Barnsbury | 230 | 37.6174735 | 28.1886718 | 0 | 100.00 |
| Brixton | 49 | 37.2316432 | 16.6568589 | 0 | 79.1057105 |
| Dalston | 110 | 33.8434717 | 14.2990224 | 0 | 78.2605761 |

Tab. 3a. ANOVA outputs and summary statistics for Ba, Br and Da.

Ba and Th exhibit slightly different values, respectively higher and lower. The case of Ba however shows a much wider distribution, which is due to a remarkable presence of squares and unbuilt areas in the urban fabric. We present in Tab. 3b the quartile intervals of M5 and in Fig. 6 the distributions of M5 for the five case studies.

| Neighbourhood | Q1 | Q2 | Q3 |
|---|---|---|---|
| Barnsbury | 0.0000 | 38.5821 | 73.7727 |
| Battersea | 41.8441 | 50.4922 | 56.5027 |
| Brixton | 27.6311 | 39.1180 | 47.1906 |
| Dalston | 27.3630 | 33.6800 | 39.8845 |
| Telegraph Hill | 21.0771 | 25.2760 | 31.0538 |

Tab. 3b. Quartile intervals of M5 for the five case studies.

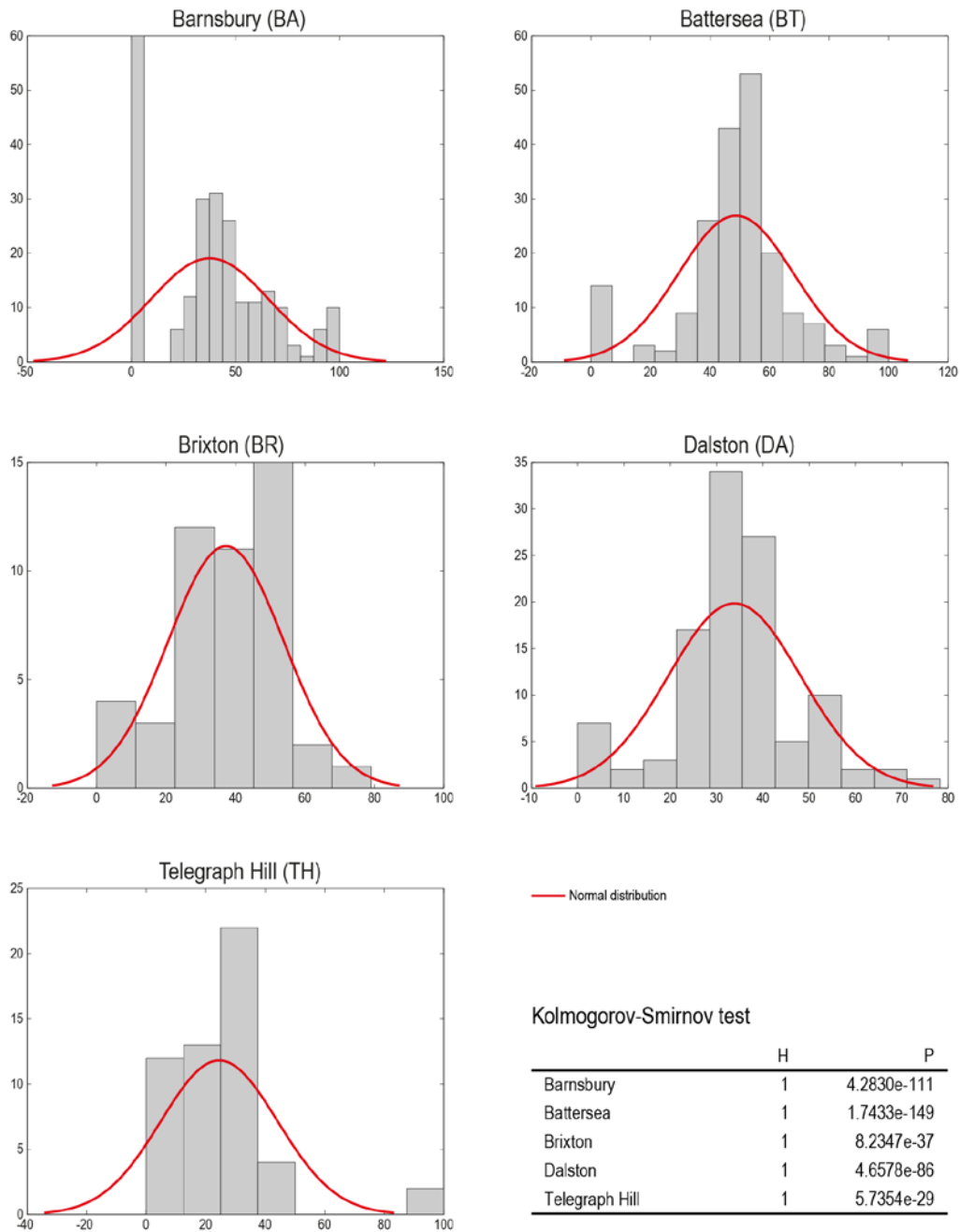

Fig. 6. Distributions of *Coverage ratio (M5)* for the five neighbourhoods.

As for *Density (M3)*, the mean values in all neighbourhoods sit around 1.0 (1m² of floor area per m² of street edge area which equals to roughly 100 units per hectare), with Th being the only exception (circa 0,6). This different behaviour seems to be linked with a bigger value of *Street edge (M1)* as previously shown.

The ANOVA procedure allows us to say that the hypothesis

$$H_0 : \mu_{Ba} = \mu_{Bt} = \mu_{Br} = \mu_{Da}$$

can be accepted. This means that, for what concerns M3, Ba, Bt, Br and Da are statistically equal. We present in Tab. 4a ANOVA outputs and summary statistics for Ba, Br and Da; in Tab. 4b, quartile intervals of M5 for all the case studies and in Fig. 7 M5 distributions.

| District | Nr. of Obs. | DF | F | Pr>F |
|---|---|---|---|---|
| Ba-Bt-Br-Da | 586 | 3 | 2.15 | 0.0933 |

| Neighbourhood | Nr. of Obs. | Mean | Std. Dev. | Min. | Max. |
|---|---|---|---|---|---|
| Barnsbury | 230 | 1.0533917 | 0.8452948 | 0 | 4.0823944 |
| Battersea | 197 | 0.9586052 | 0.4723887 | 0 | 2.2049346 |
| Brixton | 49 | 0.8210166 | 0.3563589 | 0 | 1.6215977 |
| Dalston | 110 | 0.9138698 | 0.7322159 | 0 | 4.5415537 |

Tab. 4a. ANOVA outputs and summary statistics for Ba, Br and Da.

| Neighbourhood | Q1 | Q2 | Q3 |
|---|---|---|---|
| Barnsbury | 0.0000 | 1.04902 | 1.50438 |
| Battersea | 0.826689 | 0.997633 | 1.139319 |
| Brixton | 0.685211 | 0.863607 | 1.028128 |
| Dalston | 0.632385 | 0.754816 | 1.143714 |
| Telegraph Hill | 0.437457 | 0.534967 | 0.650091 |

Tab. 4b. Quartile intervals of M5 for the five case studies.

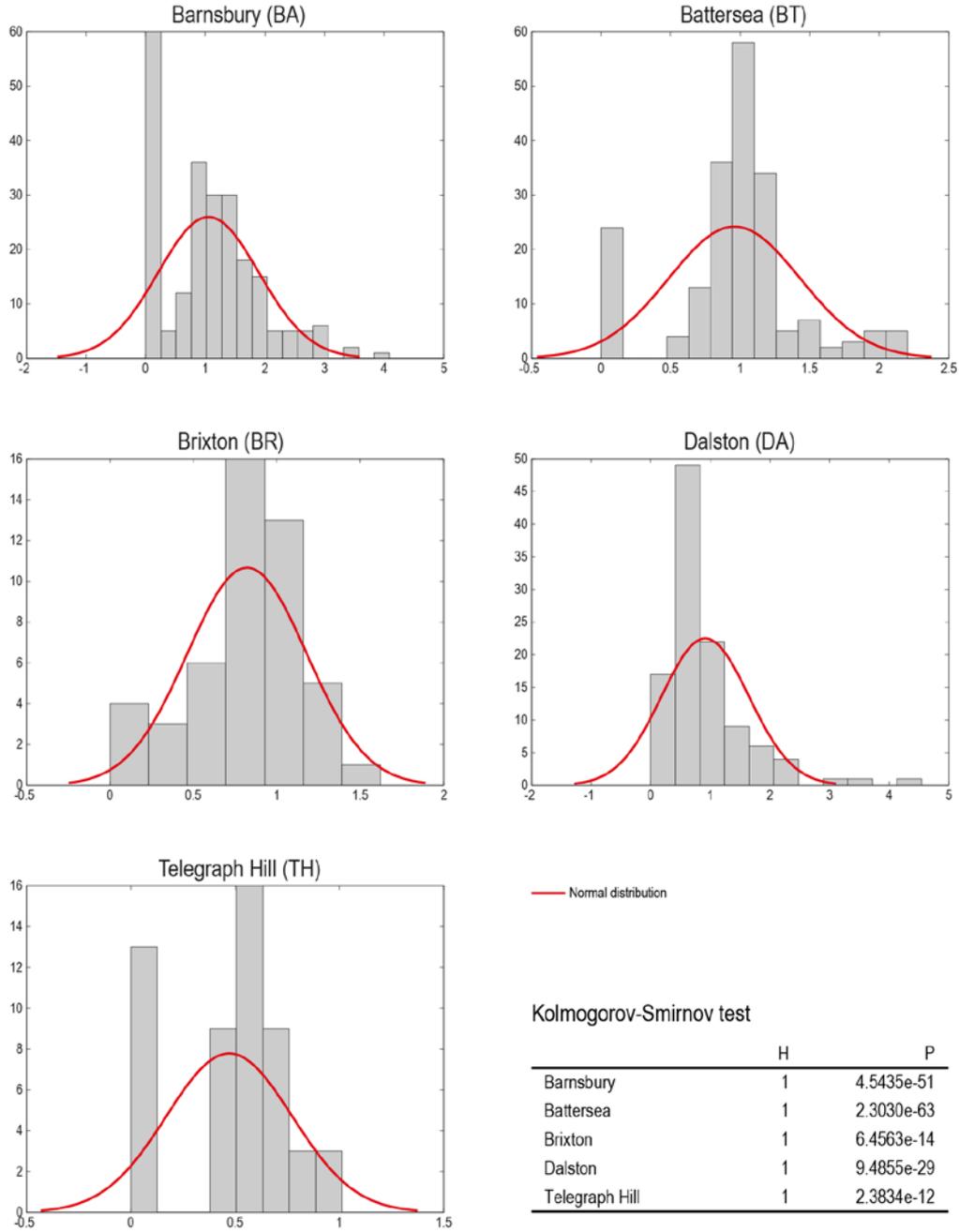

Fig. 7. Distributions of *Density (M3)* for the five neighbourhoods.

The *Streets width (M4)* sits around a value of 8m for all cases, with Bt-Br and Ba-Da as extreme cases. We show that the mean values of Bt and Br pass the hypothesis test

$$H_0 : \mu_{Bt} = \mu_{Br}$$

with the usual level of confidence. Moreover, considering the same level of confidence, Ba and Da pass the hypothesis test with an even stronger evidence

$$H_0 : \mu_{Ba} = \mu_{Da}.$$

Results of the ANOVA test for the pairs Bt-Br are shown in Tab. 5a.

| Neighbourhoods | Nr. of Obs. | DF | F | Pr>F |
|---|---|---|---|---|
| Bt-Br | 246 | 1 | 2.74 | 0.0992 |

| Neighbourhood | Nr. of Obs. | Mean | Std. Dev. | Min | Max |
|---|---|---|---|---|---|
| Battersea | 197 | 7.7403571 | 0.9723803 | 6.18 | 12.07 |
| Brixton | 49 | 8.0364583 | 1.5597551 | 6.54 | 12.35 |

Tab. 5a. ANOVA outputs and summary statistics for Ba and Br.

While ANOVA outputs for Ba-Da are presented in Tab. 5b.

| Neighbourhoods | Nr. of Obs. | DF | F | Pr>F |
|---|---|---|---|---|
| Ba-Da | 340 | 1 | 0.21 | 0.6441 |

| Neighbourhood | Nr. of Obs. | Mean | Std. Dev. | Min | Max |
|---|---|---|---|---|---|
| Ba | 230 | 8.6938261 | 2.3942543 | 2.75 | 16.57 |
| Da | 110 | 8.81 | 1.8816459 | 3.79 | 14.63 |

Tab. 5b. ANOVA outputs and summary statistics for Ba and Da.

We observe that M4 values for the five neighbourhoods range over a small interval of values with Th being the only significant exception (mean value higher than 10m). Th behaves differently also in the distributions (Fig. 8). However values of M4 for this neighbourhood remain in the same order of magnitude of the others (between 8 and 12m).

| Neighbourhood | Q1 | Q2 | Q3 |
|---|---|---|---|
| Barnsbury | 7.200 | 9.055 | 10.210 |
| Battersea | 7.200 | 7.495 | 7.900 |
| Brixton | 7.07 | 7.48 | 7.93 |
| Dalston | 8.14 | 8.79 | 9.55 |
| Telegraph Hill | 8.78 | 11.84 | 12.97 |

Tab. 5c. Quartile intervals of M4 for the five case studies.

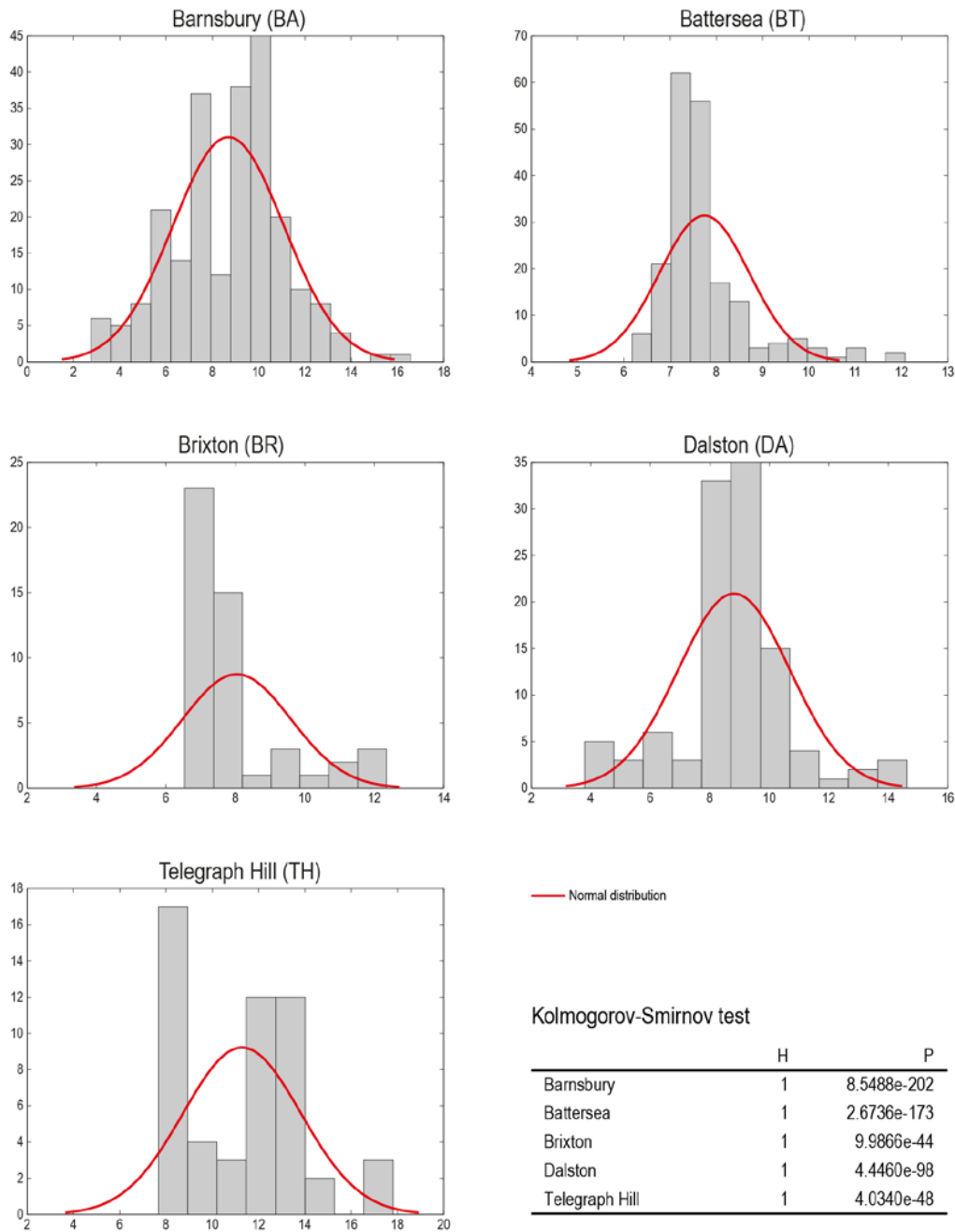

Fig. 8. Distributions of *Street width (M4)* for the five neighbourhoods.

The hypothesis test of mean values for the variable Street centrality (M6) does not have statistical relevance. This might be due to the fact that the numerical value of centrality does not have practical meaning; it measures, in fact, a degree of connectivity rather than a dimension in space. Although M6 does not pass the hypothesis test presented above, it might be useful to explain behaviours of the other variables through regression analysis. We will explore this in the next section.

As for the other two indices *Front height (M7)* and *Built front ratio (M8)*, the ANOVA analysis does not identify statistical equalities between cases (Pr>F always <0.05). However, the descriptive analysis based on boxplot (Fig. 5) shows that all neighbourhoods exhibit a quite similar mean of M7 sitting at 2.5 floors with variations within the IQR ranging between 1 to 3 floors only. Similarly, the *Built front ratio (M8)* takes mean values between the 50-70%, with Bt reaching the 80% and Th dropping to the 40%. These values are typical of a perimeter block urban type, with or without front gardens between the building and the street. We also note that median values for all cases except Th are significantly higher than the means; this demonstrates the relevant presence of outliers at the bottom scale of the values (i.e. street fronts completely unbuilt or built up at a very low rate). The urban type of gentrified neighbourhoods is therefore clearly closer to a construction along the perimeter of blocks rather than a "towers in the park" or set-back type.

Finally, as further analysis on data distribution, we performed the Komogorov-Smirnov test. We observe that *H* is equal to 1 for every performed test. This means that any empirical distribution function does not come from a normal distribution. Moreover *P-values* for each variable take values near 0; this indicates that our estimation is highly precise. Results of this test are given with the analysis of data distributions in Fig. from 4 to 8.

In Fig. 9 we summarize findings for all variables with a boxplot analysis of variance. The boxplot reports the following information: third quartile (top edge of the box), median (black solid line), mean (black cross), first quartile (bottom edge of the box), minimum and maximum of data (top and bottom ends of the vertical dashed line).

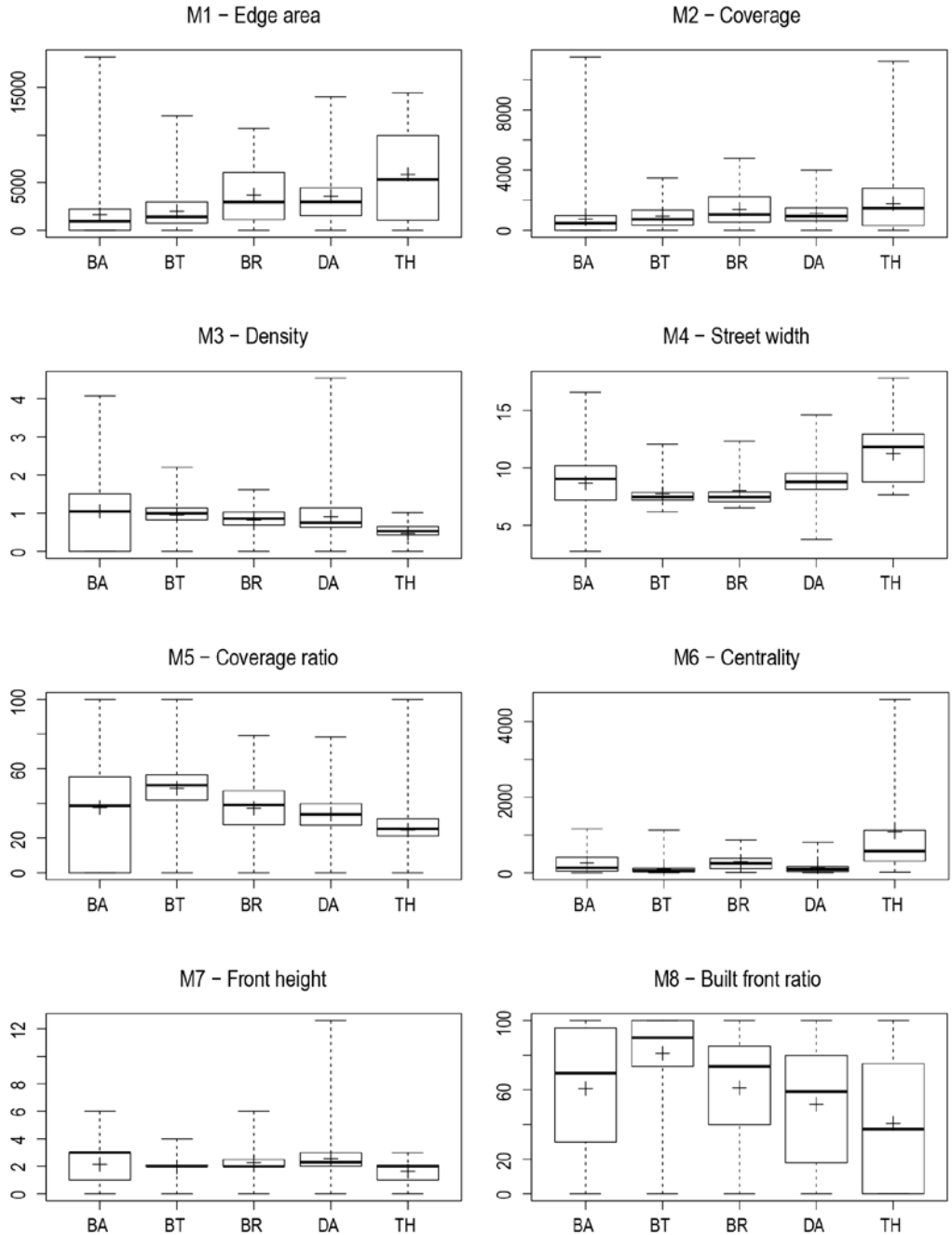

Fig. 9. Boxplots of all eight indices of urban fabric compared across the five case studies.

3.2.2. Analysis of correlation

Significant linear correlations have been found across many variables; however recurrences across all or most cases have been detected in few cases only. In particular (Tab. 6), it seems that the higher the *Density (M3)*, higher the *Coverage Ratio (M5)* with Dalston taking a correlation value slightly lower than the threshold considered but still showing a moderate value (ρ=0.47); this outcome excludes the presence of "tower blocks" type, which usually comes with large parcels of land often coinciding with the block itself and presents a much lower building coverage ratio. If this urban type was significantly present in any of our cases, we would have found that, for larger street edges, the amount of covered land would not have been significantly larger, and certainly not showing a linear correlation. For the same reason, if there was a significant presence of "tower block" types in our cases, we would have noticed that denser street edges would not have been accompanied by a larger coverage ratio, which in fact is what we observe. In short, this finding confirms that all or most of the neighbourhoods selected share the same type of urban form, i.e. a traditional low/medium rise, perimeter block.

It is then worth noting that *Centrality (M6)* does not appear to be significantly correlated with any other variable (the only exception being *Street width (M4)* in Brixton). If it did not significantly contribute to explain the variance of other variables either, street centrality would actually have played a very modest role in driving the urban form altogether (i.e. it would be an isolated factor). As we will show in the next section, this is not the case. The lack of correlation, in this case, reinforces the role of street centrality as an independent driver of urban form.

| | Density (M3) and Coverage Ratio (M5) | Density (M3) and Built front ratio (M8) | Coverage ratio (M5) and Front height (M7) | Edge area (M1) and Front height (M7) | Coverage ratio (M5) and Built front ratio (M8) | Front height (M7) and Built front ratio (M8) | Street width (M4) and Centrality (M6) | Street width (M4) and Built front ratio (M8) | Edge area (M1) and Density (M3) |
|---|---|---|---|---|---|---|---|---|---|
| Barnsbury | 0,92 | 0,52 | 0,67 | -- | 0,51 | 0,62 | -- | -- | -- |
| Battersea | 0,54 | -- | -- | -- | -- | -- | -- | -- | -- |
| Brixton | 0,90 | 0,52 | -- | -- | 0,58 | -- | 0,70 | -0,56 | -- |
| Dalston | -- | -- | -- | -- | -- | -- | -- | -- | -- |
| Telegraph Hill | 0,60 | -- | -- | 0,62 | -- | -- | -- | -- | 0,58 |

Table 6. Cases of significant correlation ($\rho$>0.5, with *P-Value* < 0.0001) among variables across the five neighbourhoods.

### 3.2.3. Regression analysis

In Tab. 7 we present the linear regression analysis outcomes for the eight variables, calculated on the unified dataset of all case studies. We note that the *Coverage ratio (M5)* contribution to explain *Density (M3)* is indeed very high ($R^2$=0.84) and this confirms the findings obtained through correlation analysis highlighted in the previous section.

Secondly, the linear model that takes *Centrality (M6)* as dependent variable gives the weakest result ($R^2$=0.35). However, *Street width (M4)*, *Built front ratio (M8)* and *Coverage (M2)* altogether describe almost the whole of it. We notice that *Street width (M4)* shows a similar behaviour with the main contributor being *Front height (M7)*. These results seem to suggest a link between features of city form and properties of streets. We thus report in Tab. 8 the linear regression for the sole *Street centrality (M6)* for each neighbourhood taken separately. Firstly, we notice that the models generally give a good explanation for the variability of *Centrality (M6)* and this is particularly noticeable for Ba, Br and Th. Moreover we observe that the main contribution to the explanation of *Street centrality (M6)* comes almost invariably from *Street width (M4)*.

These findings confirm that both the front heights and the centrality of streets contribute significantly to explain the width of the streets. This again appears to be a typical character of traditional urban form, where the urban significance of streets is linked to such formal characters. Moreover, they suggest that, within the evident limits of this study, the combination of these essential features makes the urban fabric flexible enough to respond to the subtle complexity of emerging processes of gentrification in time.

| M1 | Edge area | | | | |
|---|---|---|---|---|---|
| | Variable | Partial $R^2$ | $R^2$ | F | Pr > F |
| 1 | M2 Coverage | 0,8267 | 0,8267 | 3659,47 | < 0,0001 |
| 2 | M4 Street width | 0,0075 | 0,8343 | 34,87 | < 0,0001 |
| 3 | M5 Coverage ratio | 0,0185 | 0,85,27 | 95,89 | < 0,0001 |
| 4 | M7 Front height | 0,0065 | 0,8593 | 35,42 | < 0,0001 |
| 5 | M3 Density | 0,0123 | 0,8715 | 72,87 | < 0,0001 |
| 6 | M8 Built front ratio | 0,0003 | 0,8718 | 1,68 | 0,1954 |
| 7 | M6 Centrality | 0,0001 | 0,8719 | 0,58 | 0,4482 |

| M2 | Coverage | | | | |
|---|---|---|---|---|---|
| | Variable | Partial $R^2$ | $R^2$ | F | Pr > F |
| 1 | M1 Edge area | 0,8267 | 0,8267 | 3659,47 | < 0,0001 |
| 2 | M5 Coverage ratio | 0,0277 | 0,8544 | 145,69 | < 0,0001 |
| 3 | M7 Front height | 0,0051 | 0,8595 | 27,90 | < 0,0001 |
| 4 | M3 Density | 0,0087 | 0,8682 | 50,43 | < 0,0001 |
| 5 | M6 Centrality | 0,0008 | 0,8690 | 4,67 | 0,0311 |
| 6 | M4 Street width | 0,0008 | 0,8698 | 4,51 | 0,034 |
| 7 | M8 Built front ratio | 0,0006 | 0,8704 | 3,76 | 0,0527 |

| M3 | Density | | | | |
|---|---|---|---|---|---|
| | Variable | Partial $R^2$ | $R^2$ | F | Pr > F |
| 1 | M5 Coverage ratio | 0,8427 | 0,8427 | 4107,86 | < 0,0001 |
| 2 | M7 Front height | 0,0737 | 0,9164 | 675,13 | < 0,0001 |
| 3 | M4 Street width | 0,0110 | 0,9274 | 116,31 | < 0,0001 |
| 4 | M1 Edge area | 0,0030 | 0,9304 | 32,97 | < 0,0001 |
| 5 | M2 Coverage | 0,0036 | 0,9340 | 41,30 | < 0,0001 |
| 6 | M6 Centrality | 0,0002 | 0,9342 | 2,44 | 0,1183 |
| 7 | M8 Built front ratio | 0,0001 | 0,9342 | 0,62 | 0,4311 |

| M4 | Street width | | | | |
|---|---|---|---|---|---|
| | Variable | Partial $R^2$ | $R^2$ | F | Pr > F |
| 1 | M7 Front height | 0,7040 | 0,7040 | 1824,08 | < 0,0001 |
| 2 | M6 Centrality | 0,0481 | 0,7521 | 148,49 | < 0,0001 |
| 3 | M8 Built front ratio | 0,0428 | 0,7948 | 159,49 | < 0,0001 |
| 4 | M5 Coverage ratio | 0,0115 | 0,8063 | 45,20 | < 0,0001 |
| 5 | M3 Density | 0,0220 | 0,8283 | 97,93 | < 0,0001 |
| 6 | M1 Edge area | 0,0045 | 0,8328 | 20,63 | < 0,0001 |
| 7 | M2 Coverage | 0,0014 | 0,8343 | 6,55 | 0,0107 |

| M5 | Coverage ratio | | | | |
|---|---|---|---|---|---|
| | Variable | Partial $R^2$ | $R^2$ | F | Pr > F |
| 1 | M3 Density | 0,8427 | 0,8427 | 4107,86 | < 0,0001 |
| 2 | M4 Street width | 0,0275 | 0,8702 | 162,29 | < 0,0001 |
| 3 | M7 Front height | 0,0140 | 0,8842 | 92,40 | < 0,0001 |
| 4 | M8 Built front ratio | 0,0087 | 0,8929 | 62,34 | < 0,0001 |
| 5 | M2 Coverage | 0,0022 | 0,8951 | 16,11 | < 0,0001 |
| 6 | M6 Centrality | 0,0004 | 0,8955 | 2,79 | 0,0954 |
| 7 | M1 Edge area | 0,0003 | 0,8958 | 1,97 | 0,1609 |

| M6 | Centrality | | | | |
|---|---|---|---|---|---|
| | Variable | Partial $R^2$ | $R^2$ | F | Pr > F |
| 1 | M4 Street width | 0,2887 | 0,2887 | 311,28 | < 0,0001 |
| 2 | M8 Built front ratio | 0,0326 | 0,3212 | 36,75 | < 0,0001 |
| 3 | M2 Coverage | 0,0169 | 0,3381 | 19,48 | < 0,0001 |
| 4 | M7 Front height | 0,0041 | 0,3422 | 4,81 | 0,0286 |
| 5 | M1 Edge area | 0,0007 | 0,3430 | 0,84 | 0,3599 |
| 6 | M5 Coverage ratio | 0,0009 | 0,3439 | 1,08 | 0,2988 |
| 7 | M3 Density | 0,0017 | 0,3456 | 1,93 | 0,1655 |

| M7 | Front height | | | | |
|---|---|---|---|---|---|
| | Variable | Partial $R^2$ | $R^2$ | F | Pr > F |
| 1 | M3 Density | 0,8369 | 0,8369 | 3936,81 | < 0,0001 |
| 2 | M4 Street width | 0,0425 | 0,8795 | 270,37 | < 0,0001 |
| 3 | M5 Coverage ratio | 0,0130 | 0,8925 | 92,40 | < 0,0001 |
| 4 | M1 Edge area | 0,0084 | 0,9009 | 64,81 | < 0,0001 |
| 5 | M2 Coverage | 0,0064 | 0,9073 | 52,53 | < 0,0001 |
| 6 | M8 Built front ratio | 0,0008 | 0,9081 | 6,90 | 0,0088 |
| 7 | M6 Centrality | 0,0005 | 0,9086 | 4,08 | 0,0438 |

| M8 | Built front ratio | | | | |
|---|---|---|---|---|---|
| | Variable | Partial $R^2$ | $R^2$ | F | Pr > F |
| 1 | M5 Coverage ratio | 0,6187 | 0,6187 | 1244,59 | < 0,0001 |
| 2 | M4 Street width | 0,0394 | 0,6582 | 88,39 | < 0,0001 |
| 3 | M6 Centrality | 0,0116 | 0,6698 | 26,91 | < 0,0001 |
| 4 | M7 Front height | 0,0025 | 0,6723 | 5,87 | 0,0156 |
| 5 | M2 Coverage | 0,0009 | 0,6732 | 2,13 | 0,1449 |
| 6 | M1 Edge area | 0,0007 | 0,6739 | 1,53 | 0,2166 |
| 7 | M3 Density | 0,0003 | 0,6741 | 0,62 | 0,4311 |

Tab. 7. Linear regression analysis for the eight indices calculated on data from all five neighbourhoods.

| Ba | Barnsbury | | | | |
|---|---|---|---|---|---|
| | Variable | Partial $R^2$ | $R^2$ | F | Pr > F |
| 1 | M4 Street width | 0,5037 | 0,5037 | 232.44 | < 0,0001 |
| 2 | M3 Density | 0,0463 | 0,5500 | 23,46 | < 0,0001 |
| 3 | M2 Coverage | 0,0109 | 0,5609 | 5,64 | 0,0184 |
| 4 | M8 Built front ratio | 0,0115 | 0,5724 | 6,06 | 0,0146 |
| 5 | M1 Edge area | 0,0019 | 0,5743 | 1,02 | 0,3130 |

| Bt | Battersea | | | | |
|---|---|---|---|---|---|
| | Variable | Partial $R^2$ | $R^2$ | F | Pr > F |
| 1 | M4 Street width | 0,3663 | 0,3663 | 112,72 | < 0,0001 |
| 2 | M7 Front height | 0,0190 | 0,3853 | 5,99 | 0,0152 |
| 3 | M5 Coverage ratio | 0,0103 | 0,3956 | 3,28 | 0,0717 |
| 4 | M1 Edge area | 0,0025 | 0,3981 | 0,81 | 0,3691 |
| 5 | M3 Density | 0,0015 | 0,3996 | 0,49 | 0,4844 |
| 6 | M2 Coverage | 0,0021 | 0,4018 | 0,67 | 0,4144 |

| Br | Brixton | | | | |
|---|---|---|---|---|---|
| | Variable | Partial $R^2$ | $R^2$ | F | Pr > F |
| 1 | M4 Street width | 0,7280 | 0,7280 | 125,79 | < 0,0001 |
| 2 | M3 Density | 0,0226 | 0,7506 | 4,17 | 0,0469 |

| Da | Dalston | | | | |
|---|---|---|---|---|---|
| | Variable | Partial $R^2$ | $R^2$ | F | Pr > F |
| 1 | M4 Street width | 0,4323 | 0,4323 | 83,00 | < 0,0001 |
| 2 | M8 Built front ratio | 0,0283 | 0,4606 | 5,68 | 0,0189 |
| 3 | M3 Density | 0,0044 | 0,4651 | 0,89 | 0,3482 |
| 4 | M2 Coverage | 0,0032 | 0,4682 | 0,63 | 0,4291 |

| TH | Telegraph Hill | | | | |
|---|---|---|---|---|---|
| | Variable | Partial $R^2$ | $R^2$ | F | Pr > F |
| 1 | M3 Density | 0,5274 | 0,5274 | 58,02 | < 0,0001 |
| 2 | M1 Edge area | 0,0656 | 0,5930 | 8,22 | 0,0060 |
| 3 | M4 Street width | 0,0559 | 0,6488 | 7,95 | 0,0069 |
| 4 | M8 Built front ratio | 0,0466 | 0,6954 | 7,49 | 0,0086 |
| 5 | M7 Front height | 0,0275 | 0,7229 | 4,76 | 0,0340 |
| 6 | M2 Coverage | 0,0086 | 0,7314 | 1,50 | 0,2271 |

Tab. 8. Linear regressions of *Centrality (M6)* for the five gentrified areas taken separately.

### 3.3. A narrative of the typical gentrified area

We can summarize the results of this quantitative investigation of five gentrified urban neighbourhoods in London in a narrative that links the common patterns emerging from an analysis of their forms.

The correlation analysis shows that street *Centrality (M6)* is not correlated with any of the variables overall. However, on a case by case basis, this becomes significantly explained by most other variables, especially by *Street width (M4)* and, at lower grade, *Density (M3)*. This reinforces the idea that extremely complex dynamics occurring at different scales between mutually interrelated historical factors in time, which produce the form of a city, are related to the connectivity of places as determined by the way streets are linked up with one another in space.

In particular, by visually analysing the geographic location of centrality, the typical gentrified urban area sits between streets of very high centrality, which constitute its boundaries. However, main streets of a lower grade ("local mains") often traverse the area, Barnsbury (Liverpool Road) and Telegraph Hill (Pepys Road) being clear examples. Differently to the highly central streets at their boundaries, the traversing local mains rarely attract consistent retail commerce uses, with the only exceptions of local businesses (cafes, newsagents, groceries) especially emerging at crossings with highly central streets. This confirms the double nature of good urban districts as outlined by Jacobs and Appleyard (A. Jacobs & Appleyard, 1987) where they advocate for "*s*anctuary areas" that are *"well-managed environment relatively devoid of nuisance, overcrowding, noise, danger, air pollution, dirt, trash, and other unwelcome intrusions"*, which would nevertheless be within easy reach from places where people *"can* *break from traditional molds, extend their experience, meet new people, learn other viewpoints, have fun"* (A. Jacobs & Appleyard, 1987, p. 116).

The highly central streets at the boundaries of the gentrified areas provide links to public transport, retail and other important non-residential uses at the urban scale at walking distance (400-500m) from anywhere in the area, while local mains serve the inner residential clusters with local services and accessible routes located reasonably at hand (200-250m). This result suggests that in order for an urban area to catch up with dynamics of economic and social uplift, a multi-scalar system of central streets must be in place that features relatively tranquil residential "sanctuary areas" framed by handy main streets of different level of centrality, with the most central at their borders.

Similarities found for all urban form variables, explored through the analysis of variance, show that the prevailing urban type in all cases consistently aligns with a low/medium rise, traditional perimeter block. This prevailing model, characterized by street edges sized 4-5,000m$^2$, seems to coexist with significantly different situations, where street edges much smaller or larger may present values between 1,000m$^2$ and 6-7,000m$^2$, providing a remarkable diversity in the scale of the urban fabric, which in turns offers viable solutions for a wide variety of needs and challenges. Remarkably though, our study shows that the relative "amount of buildings", (*Density, M8*), varies across this diversity of situations in a way that is highly consistent with the relative amount of space covered by buildings (*Coverage ratio, M5*). In other words, areas developed relatively "sparsely" and relatively "intensively" nevertheless are developed using buildings of roughly similar height. In this model, the area of development is occupied at a fairly high rate by buildings, which cover between the 30-50% of it, this figure being even more remarkable if we consider that street edges may occasionally include local "pocket" parks and locally undeveloped or vacant land within the street edge's boundary. The typical density of this model equals around 1m$^2$ of gross floor area per m$^2$ of street edge, which equals roughly to 100 units per hectare. This value sits in the highest section of medium density housing, correspondent to building types such as – according to Oscar Newman (1973, p.57) – row houses, garden apartments or low town houses. In this urban type, streets are never too large, with sidewalks between 7-9m, and buildings bounding the street fronts of 2-3 floors. Finally, buildings are consistently located in close proximity to the street, so that around the 50-70% of the street front is directly lined on them.

This, in short, is the typical urban fabric that gets gentrified according to the analysis conducted over, admittedly, only five neighbourhoods in London that underwent this process. The different behaviour of Th must be noted. However, this case does not present a different urban structure altogether, but rather variations on the same traditional perimeter block structure. Firstly, we notice blocks of larger size, which give room to the location of large specialist functions within them, such as the TfL (Transport for London) Bus depot on New Cross Road, or Telegraph Hill Park. Secondly, and even more importantly, the short edges of these large blocks abut on streets (Sherwin Road and Arbuthnot Road) consistently characterized by blank walls rather than developed fronts. This unusual situation is probably determined by the low centrality of those streets, which have not exerted enough "environmental pressure" to develop the deep backyards of the fronting plots. Such blank street edges appear to be an expression of a process of densification that has not reached its peak yet.

Results presented in section 3 of this paper, and in particular narratively illustrated in section 3.3, seem to confirm that indeed all the five cases here analysed do show common traits with regards to their urban form. These traits seem to reinforce the intuitive observation that the "traditional district" features a significantly higher adaptability that makes it fitter to respond to social and economic drivers of gentrification. The traditional district seems characterized by mostly residential cores of 2-3 stories perimeter in calm and safe precincts (sanctuary areas) that are located in close proximity to local main streets (200-250m) and urban main roads (400-500m).

Remarkably, the study evidences that street centrality is consistently "driven" by elements of form such as the width of streets and the density and height of street fronts. These typical features of the traditional urban model, subverted by modern city planning theory and practice, appear in this study to be inherently linked a higher adaptability of the urban form, which in turn is essential to give place to complex processes of gentrification.

We must however highlight again that the small size of the sampled data makes any conclusion tentative and subject to subsequent verification. In this sense, this paper should be considered an attempt to advanced statistical rigour into the field of urban morphology, i.e. a contribution the development of "a new science of cities" (Batty, 2013).

## 4. Conclusions

In this paper we propose a study of the urban form of five urban areas in London that have successfully undertaken a process of gentrification. In order to describe the urban form we identify a unit of analysis that we name "street edge", i.e. the area included in all the urban plots that face the same street; furthermore, we also define eight variables that are "structural" and easily measurable from remote using commonly available geographic repositories. Finally, we undertake a systematic quantitative analysis of the five cases using a range of statistical methods to seek for evidence of recurring patterns that would help in understanding the typical contribution that urban form can provide to the emergence of gentrification processes. Urban gentrification is here seen as a natural and cyclical force underpinning the evolution of cities, i.e. just the way cities work on the ground, and indeed an opportunity for policies of social equity to take place and operate, the ultimate scope of the paper is to help understanding how urban form can adapt and positively respond to it.

Outcomes from the quantitative analysis performed seem to agree with most of the qualitative studies presented throughout this work. In particular, we note that gentrified neighbourhoods seem to be well defined areas with major roads on the edges defining boundaries and calm streets in their cores. This provides, on one side, a good connection to main amenities and transport systems while, on the other, a safe and pleasant urban environment with some local businesses. These findings seem to be in line with Butler's observations about gentrified environments as places well connected to the centre without being centres (Butler, 2003) and with the "sanctuary area" theorized by Jacobs and Appleyard i.e. self-contained mainly residential area with chances for lively urban experiences (A. Jacobs & Appleyard, 1987). Furthermore, urban type for the five gentrified neighbourhoods is aligned with a low/medium rise housing typology (around 100 units per hectare) and this seems to confirm the observations carried out by Glass about the typical housing densities and typologies of gentrified neighbourhoods i.e. terraced houses, cottages, Victorian houses (Glass, 1964).